\documentclass[aps,PRB,preprint,superscriptaddress]{revtex4-1}  

\usepackage{amsmath,amssymb}
\usepackage{bbold}
\usepackage{float}
\usepackage{graphicx}
\usepackage{dcolumn}
\usepackage{bm}
\usepackage[colorlinks]{hyperref}
\usepackage{dsfont}
\usepackage[export]{adjustbox}
\usepackage{svg}

\usepackage[dvipsnames]{xcolor}
\definecolor{orange}{RGB}{255,165,0}

\begin{document}

\title{Performance optimization of cascaded traveling wave Josephson parametric amplifiers}

\author{I. Lilja}%
 \affiliation{
  QTF Centre of Excellence, Department of Applied Physics, Aalto University, P.O. Box 15100, FI-00076 AALTO, Finland
  }

\author{E. Mukhanova}%
 \affiliation{
  QTF Centre of Excellence, Department of Applied Physics, Aalto University, P.O. Box 15100, FI-00076 AALTO, Finland
  } 
  
\author{S. Khaldeev}%
 \affiliation{
  Department of Applied Physics, Aalto University, P.O. Box 15100, FI-00076 AALTO, Finland
  }

\author{I. Golokolenov}%
 \affiliation{
  Department of Applied Physics, Aalto University, P.O. Box 15100, FI-00076 AALTO, Finland
  }
  
\author{V. Vesterinen}%
  \affiliation{QTF Centre of Excellence, VTT Technical Research Centre of Finland Ltd, P.O. Box 1000, FI-02044 VTT, Finland
 } 
  
\author{P. J. Hakonen}%
 \affiliation{
  QTF Centre of Excellence, Department of Applied Physics, Aalto University, P.O. Box 15100, FI-00076 AALTO, Finland
  }
   \affiliation{
  Department of Applied Physics, Aalto University, P.O. Box 15100, FI-00076 AALTO, Finland
  }

\begin{abstract}

Traveling-wave parametric amplifiers (TWPAs) based on Josephson metamaterials provide broadband gain with near-quantum-limited added noise. Whereas long nonlinear metamaterial devices can deliver high gain, short arrays suffer less from dissipation, pump depletion, and internal standing waves which can degrade noise performance. Here, we demonstrate a cascaded TWPA architecture that combines the advantages of both approaches by employing a short (736-element), low-dissipation Superconducting Nonlinear Asymmetric Inductive eLement (SNAIL)-based TWPA as the first amplification stage, followed by a conventional long (1632-element) TWPA that provides additional gain. The resulting amplifier cascade achieves nearly 30 dB of total gain over a tunable bandwidth of approximately 1 GHz while maintaining added noise close to the quantum limit. Our results establish a cascade of TWPAs as a practical approach for high-gain, broadband, quantum amplification with applications in quantum information processing, multi-mode entanglement, and quantum sensing applications at microwave frequencies.

\end{abstract}

\maketitle

\subsection*{Background}

In recent years, dissipation engineering \cite{metelmann2022parametric} and low-loss superconducting metamaterials \cite{aumentado2020superconducting,esposito2021perspective} have emerged as powerful tools for quantum information processing and microwave
quantum measurement. In particular, both kinetic-inductance and Josephson-junction-based metamaterials have been employed to realize traveling-wave parametric amplifiers (TWPAs) capable of near-quantum-limited amplification \cite{white2015,macklin2015near,Zorin2016,planat2020,Perelshtein2021,ranadive2022kerr,esposito2022observation,qiu2023broadband,ranadive2025isolator,kow2026}. Kinetic-inductance-based TWPAs offer high dynamic range, broad bandwidth, and near-quantum-limited amplification \cite{zmuidzinas2012,vissers2016low,chaudhuri2017broadband,DPa_kipa_Grimsmo2022,giachero2024kinetic,faramarzi2024kinetic,howe2026kinetic}. Josephson-junction-based TWPAs, in contrast, provide a highly tunable nonlinear platform in which the nonlinear response can be engineered through the circuit design. In particular, Superconducting Nonlinear Asymmetric Inductive eLements (SNAILs) enable independent control of the third- and fourth-order nonlinearities of the Josephson inductance, allowing the nonlinear interaction to be tailored for either three-wave or four-wave mixing \cite{frattini20173,Sivak2019,fadavi2023three}. In parametric amplification, a pump tone couples the signal mode to an idler mode through the nonlinear interaction, so that amplification of the signal is accompanied by the generation of correlated fluctuations in the idler band. The work presented here employs SNAIL-based TWPAs developed at VTT Technical Research Centre of Finland \cite{vesterinen2025traveling}, which have demonstrated excellent performance both as near-quantum-limited amplifiers and as sources of microwave entanglement \cite{Perelshtein2021}.

As Josephson metamaterials can exhibit significant propagation losses \cite{Martinis2005}, shorter SNAIL arrays are expected to provide lower added noise than longer arrays \cite{planat2020}. However, shortening the array reduces the amplifier gain, making the noise added by the following amplification stages more significant. This naturally raises the question of whether low added noise and high gain can be achieved simultaneously by cascading two TWPAs, with the first optimized for minimum added noise and the second for high gain while maintaining good noise performance.

In our work, the first TWPA in the cascade is a shorter variant of VTT’s standard metamaterial array with 736 SNAIL units (S-TWPA), while the second is a longer, more conventional SNAIL TWPA with 1632 SNAILs (L-TWPA), denoted by superscripts S and L, respectively, when referring to them in the following. The shorter metamaterial array indeed exhibits superior performance in dissipation-sensitive applications, such as single-mode or two-mode squeezing operations~\cite{Lilja_2026_3rd}. 
The large gain in cascaded TWPA configuration will effectively reduce the amount of amplitude and phase noise caused by the next stage HEMT amplifier \cite{clerk_introduction_2010}, while properly adjusting temperature of the amplifier allows to achieve minimal low-frequency phase noise~\cite{Katia2026}.

This work demonstrates the use of TWPAs in a cascaded configuration and outlines the methods used to facilitate this operation. 
By exercising careful control of amplified frequency mixing products, we demonstrate a gain of nearly 30\;dB over a band of $\sim$1 GHz. This high gain is achieved while maintaining an added noise temperature of $\sim \frac{1}{2} \hbar \omega$ and a 1\;dB compression point of -101\;dBm, which are in line with the best key performance characteristics of TWPAs with significantly less gain, around 20\;dB \cite{LeGal2025,ranadive2025isolator}. 

\subsection*{Experimental Techniques}

\begin{figure}
    \centering
    \includegraphics[width=0.7\linewidth]{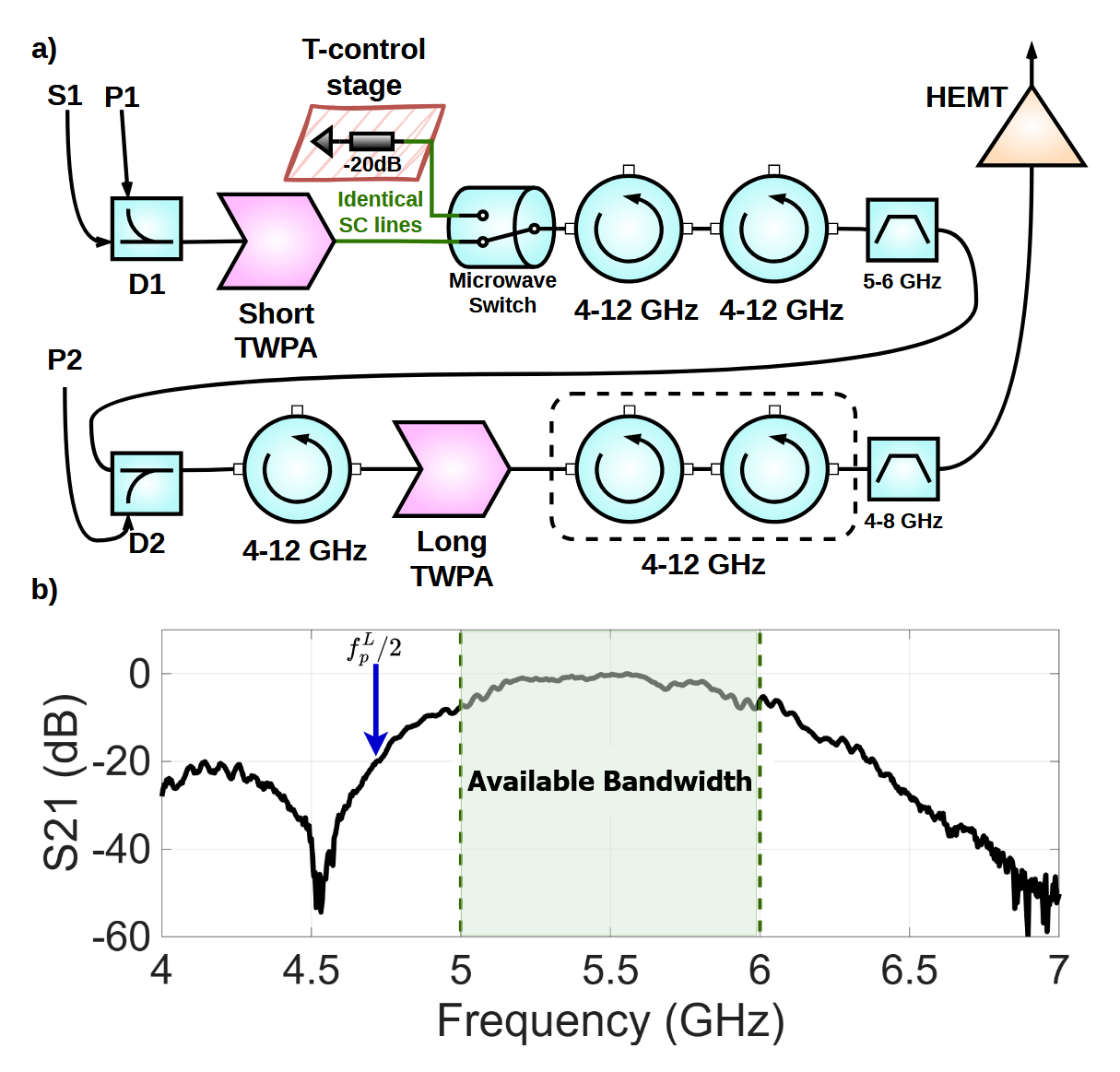}
    \caption{\textbf{Amplifier cascade: a)} Schematic of the cascaded TWPA measurement setup. The two TWPAs are driven by independent pump tones at $f^{\rm S}_{\rm p} \approx11$ GHz and $f^{\rm L}_{\rm p}\approx9.4$ GHz, inserted through their respective diplexers. An interstage bandpass filter ($5-6$\;GHz) suppresses the S-TWPA pump tone at $f^{\rm S}\sim11$\;GHz and the idler frequencies of the L-TWPA below $f < 5$\;GHz before the signal reaches the second, L-TWPA. The output band is limited by $4-8$ GHz bandpass filter and by the gain curve of the cooled LNF HEMT amplifier.  A temperature-controlled calibration stage is used for gain and noise calibration and can be selected with the microwave switch.
    \textbf{Measured transmission spectrum: b)} The pump frequency of the S-TWPA is chosen such that its signal band lies within the measurement band (5-6 GHz), whereas the second L-TWPA pump is placed outside ($f^{\rm L}_{\rm p}/2 < 5$ GHz) to suppress unwanted idler mixing products. Approximate position of half-pump frequency of the L-TWPA $f^{\rm L}_{\rm p}/2$ indicted by an arrow. }
    \label{fig:Setup}
\end{figure}

The experiment was carried out on a Bluefors dilution fridge wired for microwave measurements. Transmission characterisation at microwaves was performed using multiple devices, including Presto lock-in microwave analyzer (Intermodulation Products Ltd.), vector network analyzer (Agilent Ltd.), and heterodyning microwave signal analyzer (Anritsu Ltd.). The schematic of microwave lines is illustrated in Fig.~\ref{fig:Setup}, where only the most relevant elements are indicated.

In order to realize cascaded TWPAs, the naive approach would be to double all microwave components required to operate a single TWPA in series to each other, which includes a diplexer to add a pump tone to the signal carrier and an isolator after the TWPA to suppress back-action noise. To properly characterize the noise performance of the cascade, a thermally de-coupled temperature-controlled calibration stage was available through a microwave switch into the setup.

However, operating TWPAs in cascade presents several fundamental challenges. Most importantly, the wideband nature of these devices makes the later stage susceptible to saturation caused by quantum-level noise amplified by the preceding stage. The quantum noise at 6 GHz equals 10 pV$_{\rm{rms}}/\sqrt{\rm{Hz}}$ which over 4 GHz band yields rms-voltage noise of $\sim 1$ $\mu$V$_{\rm{rms}}$, which corresponds to -107 dBm of power. Typical input referred compression points of TWPAs are in the range $-110 $ to $ -100$ dBm for most devices when the wideband gain is around 20~dB  \cite{LeGal2025,ranadive2025isolator,Nilsson2023theory,Peng2022}. Hence, in such naive series configuration, even a modest gain of 10 dB in the first stage is sufficient to saturate the second stage. 

Additionally, because the parametric interaction couples the signal and idler modes, the first S-TWPA amplifies noise in both bands, populating the idler input of the second \mbox{L-TWPA} with excess noise. Since quantum-limited operation requires a clean, effectively vacuum (``cold'') idler mode, this excess noise degrades the performance of the second stage. Filtering the idler band can restore this condition and suppress the resulting nonlinear noise transfer to the signal band, reducing the added noise relative to the unfiltered case by an amount that depends on the gain of the S-TWPA~\cite{malnou2024low}.

Furthermore, due to the process of 3-wave mixing taking place in the first stage of the cascade, the original signal produces a strong idler. 
It effectively becomes an extra signal for the second amplification stage and creates another idler of its own. 
As a result, each input signal typically gives rise to three idler tones, significantly increasing spectral complexity leading to problems in the signal interpretation along the amplification chain. Additional spectral components may  arise from intermodulation between the various signals and idlers already present in the system \cite{remm2023intermodulation}.

All these issues can be addressed simultaneously by implementing appropriate filtering and proper adjustment of the frequency of the pump tones (see Fig.~\ref{fig:Setup}). 
A bandpass filter (5–6 GHz in our setup) reduces the bandwidth over which the amplified quantum noise from the preceding stage can reach the following stage, thus reducing the total noise power injected into the latter. In addition, the filter can be chosen so that the half-pump frequency of the L-TWPA lies outside its passband. This suppresses the thermally populated (“hot”) idler spectrum generated around the half-pump frequency and thereby prevents these excess fluctuations from participating in the parametric interaction.
Similarly, strong idlers are completely filtered out, which prevents complication of the spectrum and avoids unwanted intermodulations in the system.

In a standard setup, a 4–8/12~GHz bandpass filter is typically sufficient to limit thermal noise coming from further stages.
However, for cascaded operation this bandwidth is too broad.
In general, a narrower filter bandwidth allows the cascade to reach a higher overall gain before reaching the 1 dB compression point by reducing the extraneous noise and spurious tones passed on to the subsequent stage.

Incorporation of a bandpass filter in the cryogenic environment comes at a cost of imperfect impedance matching between subsequent amplification stages and  leads to a rise of standing waves. Poor impedance matching causes stronger reflections, which in turn produce ripples in the gain spectrum of the TWPA \cite{kern2023reflection,peng2022floquet}, which is why an additional circulator is placed after the bandpass filter to better enforce a $50\;\Omega$ environment over our bandwidth.  
This significantly mitigated the ripples in the working bandwidth, even though they still remain noticeable, particularly at L-TWPA gain exceeding 15~dB. 

The last element of the microwave setup, which is fully shown in Fig.~\ref{fig:Setup}, is the PID-controlled stage with a heated 50\;$\Omega$ resistor that can be selected by a microwave switch and used as a noise source with known temperature-dependent spectral power \cite{Simbierowicz2021}. It is required for accurate calibration of the gain using the Y-factor method \cite{Pozar}.
The cable connecting the calibration stage was chosen to be identical to the cable connected to the S-TWPA output, ensuring that the calibration reference plane coincides as closely as possible with the input of the S-TWPA.

First in our experiments, we identify a bias point for the second amplifier, L-TWPA that provides a relatively flat gain profile ($\pm 2$~dB) and a reasonably low noise temperature ($T^{\rm L}_{\rm Sys}\simeq 1$~K). Since we aim to achieve a gain of about 10 dB of the S-TWPA in the first stage, this value of noise temperature $T^{\rm L}_{\rm Sys}$ is sufficient, as it would contribute only by 100\;mK to the total system noise $T^{\rm Tot}_{\rm Sys}$. 
Next, we power up the S-TWPA, tuning its bias to achieve the targeted gain. To verify proper operation, we check that the system is free of gain compression and characterize the noise performance of the full cascade. This is done with a spectrum analyzer by measuring the improvement in the signal-to-noise ratio $\Delta\textrm{SNR}$ and the signal gain $G$. 

\subsection*{Results}

We have measured the gain and added noise level of the cascaded TWPA configuration and compared the performance to a single TWPA setup.
Even though separate pump tones of the two amplification stages are filtered out, their placement is nevertheless chosen to be such that they would not cross-influence each other across subsequent stages. This is important because we want to have total control of the independent pump tones and to fully avoid any influence of the former stage's pump on the later stage and vice versa.

\begin{figure}[h]
   \includegraphics[width=0.8\linewidth]{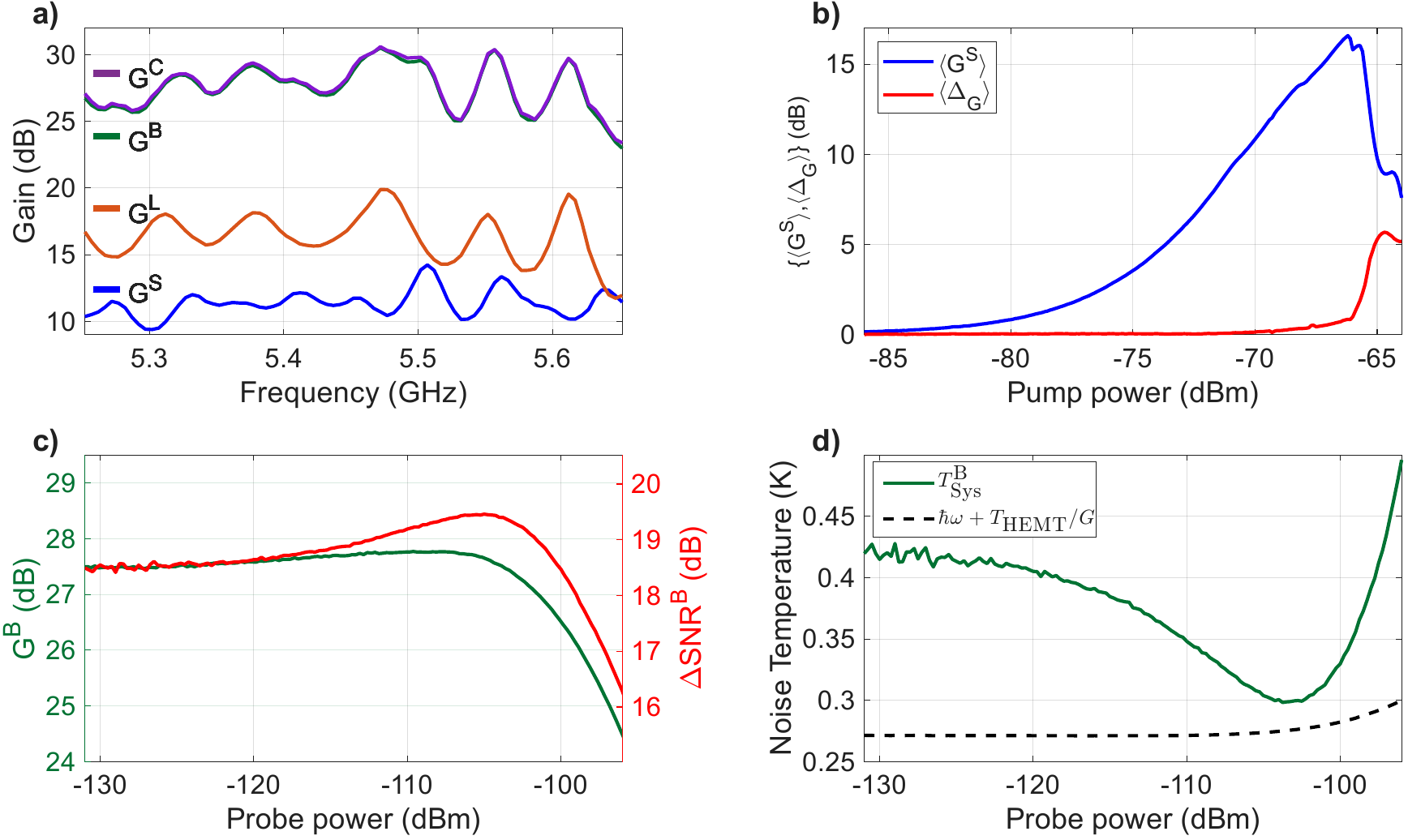}
   \caption{\textbf{a)\;Gain and compression comparison:} Gain of S-TWPA $G^\textrm{S}$ (blue trace), L-TWPA $G^\textrm{L}$ (orange trace) measured individually, simultaneously $G^\textrm{B}$ (green trace) and also plotted as linear combination of both individual measurements $G^\textrm{C}=G^\textrm{S}+G^\textrm{L}$ (purple trace) as a function of signal frequency $f_s$ around the half frequency of the S-TWPA's pump at $f^{\rm S}_p = 11$\;GHz. \textbf{b)\;Gain saturation in the cascade:} Averaged S-TWPA gain $\langle G^{S}\rangle$ (over $f=5.25 - 5.65$\;GHz) and difference between linear combination and simultaneous operation $\langle \Delta_G \rangle=\langle{}G^\textrm{C}-G^\textrm{B} \rangle$ as a function of pump power of the S-TWPA, with L-TWPA at a fixed setpoint. \textbf{c)\;Gain compression and $\Delta{}$SNR:} Measured gain $G^\textrm{B}$ (green, left axis) and corresponding improvement in signal-to-noise ratio, $\Delta \textrm{SNR}^\textrm{B}$ (red, right axis) as function of the probe power at frequency $f_{\textrm{probe}}=5.503~\textrm{GHz}$. \textbf{d)\;System noise temperature from $\Delta{}\textrm{SNR}$:} Measured total system noise temperature at frequency $f_{\textrm{probe}}=5.503~\textrm{GHz}$ with both TWPAs on as function of probe power (green line), compared to minimum achievable noise $\hbar\omega_s+T_{\textrm{HEMT}}/G^\textrm{B}$ (dashed line).}
    \label{fig:gain}
\end{figure}

The gain of the separately measured short $G^{\rm S}(f)$ (blue) and long $G^{\rm L}(f)$ (orange) TWPA is shown in Fig.~\ref{fig:gain}\;a), along with the total gain $G^{\rm B}(f)$ (green) in the cascaded configuration and the linear combination of individual gains $G^{\rm C}=[G^{\rm S}+G^{\rm L}](f)$(purple). On average, in the measured frequency range, the total observed gain overlaps with the linear combination with a precision of 0.1~dB.
The gain of the first amplifier, S-TWPA was set at 12 dB on average with maximum excursions of $\pm2$\;dB over the frequency range $5.25 \dots 5.65$\;GHz. The substantial standing-wave-like variation in the S-TWPA gain is attributed to impedance mismatch due to the presence of the bandpass filter and diplexer in between the TWPA amplifiers. 
A similar variation is observed in the gain of the L-TWPA $G^{\rm L}=17 \pm 3$\;dB. In regular operation, similar devices display a gain variation of $\pm 3$ dB for $15-20$\;dB gain \cite{Perelshtein2021}.

The maximum gain for S-TWPA alone was found to reach $\sim 15$\;dB. However, such a high gain could not be used in cascade operation because the amplified broadband noise from the first stage pushed the second amplifier, L-TWPA to its compression regime.
The gain compression of the cascaded amplifier was further investigated as a function of pump power, as demonstrated on Fig. \ref{fig:gain}\;b). There, the difference in gain $\Delta_G(f)=G^{\rm B}(f)-G^{\rm C}(f)$ between the cascaded operation  and the sum of individual gains of the TWPA devices could be seen. 

From Fig.~\ref{fig:gain}\;b) it is clear that the linear combination of individual gains matches well the observed total gain until it explodes at -66~dBm S-TWPA pump power. The difference stays below 0.1~dB up to -70~dBm pump power and below 1~dB up to -66~dBm pump power, which corresponds to the S-TWPA gain of 10 and 15~dB, respectively.

It is important to note that a change in $\Delta_G(f)$ does not necessarily indicate signal compression at the output of the L-TWPA \cite{LeGal2025}. 
Instead, simultaneous operation can alter impedance levels or phase-matching conditions, which may degrade signal transmission through the amplifier system, or in rare situations improve it. 
Alternatively, although significant effort has been made to isolate the pump signals from each other, some residual coupling, e.g. through the ground, could influence the system. 
Activating the second TWPA pump may therefore induce a slight shift in the bias point, resulting in a change in gain levels depending on the robustness of the biasing itself. 
For this reason, the device's bias should be chosen in such a way that small changes to it do not cause significant changes to the system's operation.
This problem could be alleviated by appropriate filtering between amplifiers. Depending on the application, either this issue is insignificant (typical operation for qubit's readout) or crucial (determination of the level of vacuum squeezing \cite{Perelshtein2021,Petrovnin2022} or optomechanical applications \cite{collin2022mesoscopic}).

To check the compression performance of the cascaded operation, we fixed the average gain of the S-TWPA at $G^{\rm S} \simeq 12$ dB and measured the signal transmission at fixed frequency $f_{\textrm{probe}}=5.503~\textrm{GHz}$, as shown in Fig. \ref{fig:gain}\;c) and d). There, one could see that the gain $G^S$ initially increases with an increase in the probe signal power $P_{\rm probe}$ but eventually drops at approximately -100~dBm (input-referred power). 
Such a small initial increase of gain with probe power is commonly observed in the characteristics of TWPAs and in our case remains relatively low (less than 0.5~dB) compared to fully aluminum devices ($\sim$1~dB \cite{delattre2025}). 

However, gain is not the sole figure of merit of an amplifier. Another important parameter is the signal to noise ratio (SNR), which takes into account changes in the noise level added by the device \cite{caves1982quantum}. Up to probe powers around 3~dB above the power at which the gain has already begun to decrease, SNR still improves relative to the gain. The power-induced enhancement in the SNR amounts to $\simeq 1$\;dB, indicating damping of the level of amplified vacuum noise within the TWPA when the probe signal grows. Consequently, we hypothesize non-reciprocal noise power transfer around the probe frequency against the noise around the pump frequency, which leads to apparent improvement in $T^{\textrm{S}}_{\textrm{Sys}}$ (obtained using Eq. (\ref{Eq:Addednoise}) below). The behavior of $T^{\textrm{S}}_{\textrm{Sys}}$ as a function of probe power is illustrated in Fig.~\ref{fig:gain}~d), which shows how the system noise approaches the quantum limit close to $P_{\textrm{probe}}=-103$\;dBm. Partially, this improvement can also be assigned to the saturation of the two-level systems present in the device medium at the strong applied monochromatic tone frequencies \cite{muller2019}. 

In general, the quality of an amplifier is governed by its noise characteristics that limit the improvement in SNR with increasing gain. The noise is usually summarized using a single parameter, which is the system noise temperature $T_\textrm{Sys}$.
It is given by the Friis formula:
\begin{equation}
        T_\textrm{Sys} = T_\textrm{Add}^\textrm{L}+\frac{T_\textrm{Add}^1}{G^\textrm{L}}+\frac{T_\textrm{Add}^2}{G^\textrm{L}G^1}+...,
    \label{Eq:Systemnoise}
\end{equation}
where $T_\textrm{Add}^L$ and  $G^L$ are the extra noise added by the first amplifier and its gain, and  $T_\textrm{Add}^1$, $G^1$, etc. are the parameters describing the following stages of the amplifier chain. They can be lumped into noise temperature $T_\textrm{Sys}^\textrm{L}$ that describes the system without the first stage S-TWPA. Hence, the noise temperature of the S-TWPA can be obtained from \cite{roy2015broadband}:
\begin{equation}
 T_{\textrm{Add}}^\textrm{S}= T_{\rm Sys}^\textrm{L}\left(\frac{1}{\Delta\textrm{SNR}^\textrm{S}}-\frac{1}{G^\textrm{S}}\right)-T_Q\left(1-\frac{1}{\Delta\textrm{SNR}^\textrm{S}}\right),
 \label{Eq:Addednoise}
\end{equation}
where $\Delta$SNR$^\textrm{S}$ is a change in Signal-to-Noise Ratio due to the contribution of S-TWPA, and $T_Q$ is the effective temperature of quantum fluctuations.
This noise temperature $T_\textrm{Add}^\textrm{S}$ can be divided into the unavoidable quantum part $T_Q=\frac{1}{2}\hbar \omega$ and extraneous noise $T_\textrm{Add}^\textrm{S} - T_Q$ that specifies the quality of the quantum amplifier. 
\begin{figure}[h]
    \centering
    \includegraphics[width=1\linewidth]{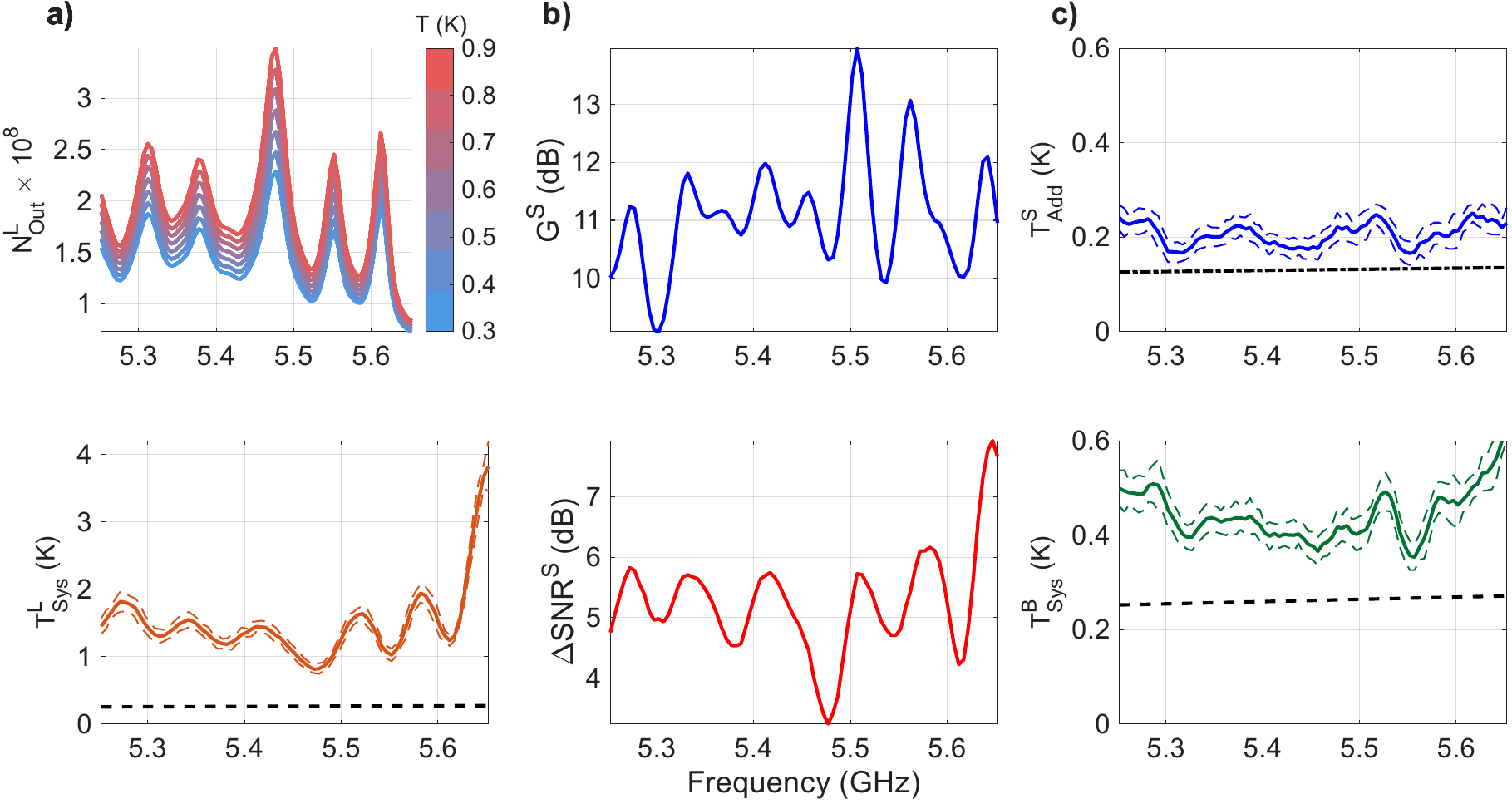}
    \caption{\textbf{a)\;Y-factor calibration:}  Top: Measured output noise of the L-TWPA as a function of the calibration-stage temperature, expressed in units of photon number (photon flux/unit band). The temperature is swept from 300 to 900 mK, with the traces color-coded according to temperature, ranging from blue (lowest temperature) to red (highest temperature). Bottom: Fitted system noise temperature obtained from the measured output noise using the Y-factor method. \textbf{b)\;S-TWPA performance:} Top: Gain of S-TWPA (zoom of the data in Fig. \ref{fig:gain}\;a). Bottom: Improvement in the signal-to-noise ratio between the S-TWPA ON and OFF states. Both quantities are obtained from spectrum analyzer measurements performed after calibrating the L-TWPA. \textbf{c)\;Noise performance:} Top: Added noise of the S-TWPA calculated using Eq. \ref{Eq:Addednoise}. Bottom: Total input-referred noise of the cascaded TWPA amplifier system. Dashed and dash-dotted lines indicate the relevant quantum limits (see main text).}
    \label{fig:NoiseTemperature}
\end{figure}
We first connect the setup to the calibration stage and perform a Y-factor calibration with only the L-TWPA powered on. This establishes the reference system noise temperature against which the contribution of the S-TWPA can be determined. The corresponding measurement is illustrated in the top of Fig.~\ref{fig:NoiseTemperature}\;a), while the bottom shows the extracted system noise temperature without the S-TWPA. 
We then determine the noise temperature of the S-TWPA from its gain and the resulting improvement in signal-to-noise ratio when it is turned on, as illustrated in Fig.~\ref{fig:NoiseTemperature}\;b).
Using these two quantities, the added noise of S-TWPA $T_\textrm{Add}^\textrm{S}$ is calculated from Eq.~\ref{Eq:Addednoise}, the result of which is shown in the top part of Fig.~\ref{fig:NoiseTemperature}\;c), while the bottom one shows the total input referred noise temperature $T_{\textrm{Sys}}^\textrm{B}$ of the cascaded TWPA amplifier.
The two noise temperatures are compared with the quantum-noise limit (black dashed lines) using different reference conventions. For the added noise of the first stage, the appropriate reference is the half-quantum limit, corresponding to the minimum added noise of a phase-preserving amplifier, given by $T_Q$. In contrast, the total input-referred system noise includes the input field's zero-point fluctuations and is therefore conventionally compared with the full quantum noise level, $\hbar\omega/k_{\mathrm B}$.

\subsection*{Conclusion}

Our work demonstrates that it is possible to combine two TWPA amplifiers in series and operate them near quantum limit with a large gain of over 30 dB. The cascade configuration, with one short and one long TWPA array of SNAILs, allows us to optimize the first stage for ultra low noise, while the second stage yields a stronger 20\;dB boost in gain. Since the first ultra-low-noise stage can provide a gain of 12\;dB, the noise contribution of the 2nd, long TWPA array remains relatively small, although its noise is not completely negligible (mean value of 115 mK)). 
In our cascaded device, the parametric pumps are placed in such a way that the noise and mixing products from the idler side of the first TWPA amplifier become completely eliminated by the inserted inter-amplifier bandpass filter. 

Our results not only demonstrate the high-quality performance of cascaded amplifiers but also suggest potential ways of improving existing TWPAs. For example, by realizing an on-chip band-pass filtering for suppression of idler frequencies or other unwanted harmonics. In another approach, a second pump tone could be injected in the middle of the SNAIL array to ``refresh'' the mixing \cite{malnou2025}. Or one could think of a combined device that would effectively realize a cascade approach fully on-chip. 

Instead of operating the two TWPAs in cascade for amplification, another option is to use the first TWPA as a source of nonclassical microwave photons\cite{Perelshtein2021,esposito2022observation,qiu2023broadband}, while the second serves as a conventional, low-noise preamplifier. Due to the intrinsically wideband nature of the TWPA, such an architecture can offer significant advantages in these types of experiments~\cite{Lilja_2026_3rd}. We expect that, in such a system, even frequency conversion microwave pumps can be operated at the quantum limit \cite{Lilian_2027}.

\subsection*{Acknowledgements}

We acknowledge the team at VTT Technical Research Centre of Finland Ltd. for
fabricating the devices: Joonas Govenius, Leif Grönberg, Robab Najafi Jabdaraghi,
Janne Lehtinen, and Mika Prunnila. This work is part of the Finnish Centre of Excellence in Quantum Materials (QMAT, RCF grant 374170). IL is grateful to the Vaisala Foundation of the Finnish Academy of Arts and Letters for a stipend. EM acknowledges a PhD scholarship from InstituteQ. SK acknowledges a PhD scholarship from QDOC. The support of the Jane and Aatos Erkko Foundation (Future Makers SELQIT project) and the Keele Foundation (SuperC project) is also gratefully acknowledged.

\bibliographystyle{unsrturl}
\bibliography{Bibliography}

@Article{zmuidzinas2012,
  author   = {Ho Eom, Byeong and Day, Peter K. and LeDuc, Henry G. and Zmuidzinas, Jonas},
  journal  = {Nature Physics},
  title    = {A wideband, low-noise superconducting amplifier with high dynamic range},
  year     = {2012},
  issn     = {1745-2481},
  number   = {8},
  pages    = {623--627},
  volume   = {8},
  doi      = {10.1038/nphys2356},
  refid    = {Ho Eom2012},
}

@article{white2015,
    author = {White, T. C. and Mutus, J. Y. and Hoi, I.-C. and Barends, R. and Campbell, B. and Chen, Yu and Chen, Z. and Chiaro, B. and Dunsworth, A. and Jeffrey, E. and Kelly, J. and Megrant, A. and Neill, C. and O'Malley, P. J. J. and Roushan, P. and Sank, D. and Vainsencher, A. and Wenner, J. and Chaudhuri, S. and Gao, J. and Martinis, John M.},
    title = {Traveling wave parametric amplifier with {J}osephson junctions using minimal resonator phase matching},
    journal = {Applied Physics Letters},
    volume = {106},
    number = {24},
    pages = {242601},
    year = {2015},
    month = {06},
    issn = {0003-6951},
    doi = {10.1063/1.4922348},
    
}

@article{kow2026,
  title = {Traveling-Wave Parametric Amplifier with Passive Reverse Isolation},
  author = {Kow, C. S. and Bell, M. T.},
  journal = {Phys. Rev. X},
  volume = {16},
  issue = {2},
  pages = {021003},
  numpages = {21},
  year = {2026},
  month = {Apr},
  publisher = {American Physical Society},
  doi = {10.1103/zkyt-r8vd},
}

@article{Simbierowicz2021,
  doi = {10.1063/5.0028951},
  year = {2021},
  month = mar,
  publisher = {{AIP} Publishing},
  volume = {92},
  number = {3},
  pages = {034708},
  author  = {Simbierowicz, Slawomir and Vesterinen, Visa and Milem, Joshua and Lintunen, Aleksi and Oksanen, Mika and Roschier, Leif and Gr{\"o}nberg, Leif and Hassel, Juha and Gunnarsson, David and Lake, Russel E.},
  title = {Characterizing cryogenic amplifiers with a matched temperature-variable noise source},
  journal = {Review of Scientific Instruments},
}

@article{Perelshtein2021,
  title = {Broadband Continuous-Variable Entanglement Generation Using a Kerr-Free {Josephson} Metamaterial},
  author = {Perelshtein, M.R. and Petrovnin, K.V. and Vesterinen, V. and Hamedani Raja, S. and Lilja, I. and Will, M. and Savin, A. and Simbierowicz, S. and Jabdaraghi, R.N. and Lehtinen, J.S. and Gr\"onberg, L. and Hassel, J. and Prunnila, M.P. and Govenius, J. and Paraoanu, G.S. and Hakonen, P.J.},
  journal = {Phys. Rev. Appl.},
  volume = {18},
  issue = {2},
  pages = {024063},
  numpages = {14},
  year = {2022},
  month = {Aug},
  publisher = {American Physical Society},
  doi = {10.1103/PhysRevApplied.18.024063},
}

@article{Petrovnin2022,
author = {Petrovnin, Kirill Viktorovich and Perelshtein, Michael Romanovich and Korkalainen, Tero and Vesterinen, Visa and Lilja, Ilari and Paraoanu, Gheorghe Sorin and Hakonen, Pertti Juhani},
title = {Generation and Structuring of Multipartite Entanglement in a {Josephson} Parametric System},
journal = {Advanced Quantum Technologies},
volume = {6},
number = {1},
pages = {2200031},
doi = {10.1002/qute.202200031},
year = {2023}
}

@article{clerk_introduction_2010,
	title = {Introduction to quantum noise, measurement, and amplification},
	volume = {82},
	issn = {0034-6861},
	doi = {10.1103/RevModPhys.82.1155},
	number = {2},
	journal = {Reviews of Modern Physics},
	author = {Clerk, A. A. and Devoret, M. H. and Girvin, S. M. and Marquardt, Florian and Schoelkopf, R. J.},
	month = apr,
	year = {2010},
	pages = {1155--1208},
}

@article{caves1982quantum,
  title={Quantum limits on noise in linear amplifiers},
  author={Caves, Carlton M},
  journal={Physical Review D},
  volume={26},
  number={8},
  pages={1817},
  year={1982},
  publisher={APS},
  doi={10.1103/PhysRevD.26.1817}
}

@article{macklin2015near,
  title={A near--quantum-limited {Josephson} traveling-wave parametric amplifier},
  author={Macklin, Chris and O’brien, K and Hover, D and Schwartz, ME and Bolkhovsky, V and Zhang, X and Oliver, WD and Siddiqi, I},
  journal={Science},
  volume={350},
  number={6258},
  pages={307--310},
  year={2015},
  publisher={American Association for the Advancement of Science},
  doi = {10.1126/science.aaa8525}
}

@article{DPa_kipa_Grimsmo2022,
  title = {Degenerate Parametric Amplification via Three-Wave Mixing Using Kinetic Inductance},
  author = {Parker, Daniel J. and Savytskyi, Mykhailo and Vine, Wyatt and Laucht, Arne and Duty, Timothy and Morello, Andrea and Grimsmo, Arne L. and Pla, Jarryd J.},
  journal = {Phys. Rev. Appl.},
  volume = {17},
  issue = {3},
  pages = {034064},
  numpages = {35},
  year = {2022},
  month = {Mar},
  publisher = {American Physical Society},
  doi = {10.1103/PhysRevApplied.17.034064},
}

@misc{vesterinen2025traveling,
  title={Traveling wave parametric amplifier},
  author={Vesterinen, Visa and Simbierowicz, Slawomir},
  year={2025},
  month=jan # "~21",
  publisher={Google Patents},
  note={US Patent 12,206,367}
}

@article{Martinis2005,
  author    = {Martinis, J. M. and Cooper, K. B. and McDermott, R. and Steffen, M. and Ansmann, M. and Osborn, K. D. and Cicak, K. and Oh, S. and Pappas, D. P. and Simmonds, R. W. and Yu, C. C.},
  title     = {Decoherence in {Josephson} Qubits from Dielectric Loss},
  journal   = {Physical Review Letters},
  volume    = {95},
  number    = {21},
  pages     = {210503},
  year      = {2005},
  doi       = {10.1103/PhysRevLett.95.210503}
}

@article{ranadive2022kerr,
  title={Kerr reversal in {Josephson} meta-material and traveling wave parametric amplification},
  author={Ranadive, Arpit and Esposito, Martina and Planat, Luca and Bonet, Edgar and Naud, C{\'e}cile and Buisson, Olivier and Guichard, Wiebke and Roch, Nicolas},
  journal={Nature Communications},
  volume={13},
  number={1},
  pages={1737},
  year={2022},
  publisher={Nature Publishing Group UK London},
  doi={10.1038/s41467-022-29375-5}
}

@book{Pozar,
      author        = "Pozar, David M",
      title         = "{Microwave engineering}",
      publisher     = "Wiley",
      address       = "Hoboken, NJ",
      year          = "2005",
      url           = "https://cds.cern.ch/record/882338",
}

@article{malnou2024low,
  title={Low-noise cryogenic microwave amplifier characterization with a calibrated noise source},
  author={Malnou, Maxime and Larson, TFQ and Teufel, JD and Lecocq, Florent and Aumentado, Joe},
  journal={Review of Scientific Instruments},
  volume={95},
  number={3},
  year={2024},
  pages={034703},
  publisher={AIP Publishing},
  doi={10.1063/5.0193591}
}

@article{Zorin2016,
  title   = {{Josephson} Traveling-Wave Parametric Amplifier with Three-Wave Mixing},
  author  = {Zorin, A.B.},
  journal = {Phys. Rev. Applied},
  volume  = {6},
  pages   = {034006},
  year    = {2016},
  doi     = {10.1103/PhysRevApplied.6.034006}
}

@article{qiu2023broadband,
  title={Broadband squeezed microwaves and amplification with a {Josephson} travelling-wave parametric amplifier},
  author={Qiu, Jack Y and Grimsmo, Arne and Peng, Kaidong and Kannan, Bharath and Lienhard, Benjamin and Sung, Youngkyu and Krantz, Philip and Bolkhovsky, Vladimir and Calusine, Greg and Kim, David and others},
  journal={Nature Physics},
  volume={19},
  number={5},
  pages={706--713},
  year={2023},
  publisher={Nature Publishing Group UK London},
  doi = {10.1038/s41567-022-01929-w}
}

@article{esposito2022observation,
  title={Observation of two-mode squeezing in a traveling wave parametric amplifier},
  author={Esposito, Martina and Ranadive, Arpit and Planat, Luca and Leger, S{\'e}bastien and Fraudet, Dorian and Jouanny, Vincent and Buisson, Olivier and Guichard, Wiebke and Naud, C{\'e}cile and Aumentado, Jos{\'e} and others},
  journal={Physical Review Letters},
  volume={128},
  number={15},
  pages={153603},
  year={2022},
  publisher={APS},
  doi = {10.1103/PhysRevLett.128.153603}
}

@article{roy2015broadband,
  title={Broadband parametric amplification with impedance engineering: Beyond the gain-bandwidth product},
  author={Roy, Tanay and Kundu, Suman and Chand, Madhavi and Vadiraj, AM and Ranadive, A and Nehra, N and Patankar, Meghan P and Aumentado, J and Clerk, AA and Vijay, R},
  journal={Applied Physics Letters},
  volume={107},
  number={26},
  year={2015},
  publisher={AIP Publishing},
  doi = {10.1063/1.4939148}
}

@article{frattini20173,
  title={3-wave mixing {Josephson} dipole element},
  author={Frattini, NE and Vool, Uri and Shankar, S and Narla, A and Sliwa, KM and Devoret, MH},
  journal={Applied Physics Letters},
  volume={110},
  number={22},
  pages={222603},
  year={2017},
  publisher={AIP Publishing},
  doi={10.1063/1.4984142}
}

@article{remm2023intermodulation,
  title={Intermodulation distortion in a {Josephson} traveling-wave parametric amplifier},
  author={Remm, Ants and Krinner, Sebastian and Lacroix, Nathan and Hellings, Christoph and Swiadek, Fran{\c{c}}ois and Norris, Graham J and Eichler, Christopher and Wallraff, Andreas},
  journal={Physical Review Applied},
  volume={20},
  number={3},
  pages={034027},
  year={2023},
  publisher={APS},
  doi={10.1103/PhysRevApplied.20.034027}
}

@article{kern2023reflection,
  title={Reflection-enhanced gain in traveling-wave parametric amplifiers},
  author={Kern, S and Neilinger, P and Il'Ichev, E and Sultanov, A and Schmelz, M and Linzen, S and Kunert, J and Oelsner, G and Stolz, R and Danilov, A and others},
  journal={Physical Review B},
  volume={107},
  number={17},
  pages={174520},
  year={2023},
  publisher={APS},
  doi={10.1103/PhysRevB.107.174520}
}

@article{Katia2026,
    title={Noise analysis in complex SQUID-based traveling wave parametric amplifiers},
    author={Mukhanova, Ekaterina and \textit{et al.}},
    journal={TBD},
    year={2026},
}

@article{Lilja_2026_3rd,
    title={Microwave cluster states with genuine multipartide entanglenment},
    author={Lilja, Ilari and \textit{et al.}},
    journal={TBD},
    year={2026},
}

@article{collin2022mesoscopic,
  title={Mesoscopic quantum thermo-mechanics: A new frontier of experimental physics},
  author={Collin, E},
  journal={AVS Quantum Science},
  volume={4},
  number={2},
  year={2022},
  pages={020501},
  publisher={AIP Publishing},
  doi={10.1116/5.0086059}
}

@article{delattre2025,
  title={Quantitative calibration of a traveling-wave parametric amplifier applied to an optomechanical platform},
  author={Delattre, Alexandre and Golokolenov, Ilya and Pedurand, Richard and Roch, Nicolas and Ranadive, Arpit and Esposito, Martina and Planat, Luca and Fefferman, Andrew and Collin, Eddy and Zhou, Xin and others},
  journal={Physical Review Applied},
  volume={24},
  number={5},
  pages={054032},
  year={2025},
  publisher={APS},
  doi={10.1103/1wgj-k7c5}
}

@article{muller2019,
  title={Towards understanding two-level-systems in amorphous solids: insights from quantum circuits},
  author={M{\"u}ller, Clemens and Cole, Jared H and Lisenfeld, J{\"u}rgen},
  journal={Reports on Progress in Physics},
  volume={82},
  number={12},
  pages={124501},
  year={2019},
  publisher={IOP Publishing},
  doi={10.1088/1361-6633/ab3a7e}
}

@article{malnou2025,
  title={A travelling-wave parametric amplifier and converter},
  author={Malnou, M and Miller, BT and Estrada, JA and Genter, K and Cicak, K and Teufel, JD and Aumentado, J and Lecocq, F},
  journal={Nature Electronics},
  volume={8},
  number={11},
  pages={1082--1088},
  year={2025},
  publisher={Nature Publishing Group UK London},
  doi={10.1038/s41928-025-01445-8}
}

@article{Lilian_2027,
    title={Simultaneous entanglement generation a´with parametric squeezing and conversion pumps},
    author={Kekkonen, Lilian and \textit{et al.}},
    journal={TBD},
    year={2026},
}

@article{planat2020,
  title = {Photonic-Crystal {Josephson} Traveling-Wave Parametric Amplifier},
  author = {Planat, Luca and Ranadive, Arpit and Dassonneville, R\'emy and Puertas Mart\'{\i}nez, Javier and L\'eger, S\'ebastien and Naud, C\'ecile and Buisson, Olivier and Hasch-Guichard, Wiebke and Basko, Denis M. and Roch, Nicolas},
  journal = {Phys. Rev. X},
  volume = {10},
  issue = {2},
  pages = {021021},
  numpages = {19},
  year = {2020},
  month = {Apr},
  publisher = {American Physical Society},
  doi = {10.1103/PhysRevX.10.021021},
}

@article{Sivak2019,
  title = {Kerr-Free Three-Wave Mixing in Superconducting Quantum Circuits},
  author = {Sivak, V.V. and Frattini, N.E. and Joshi, V.R. and Lingenfelter, A. and Shankar, S. and Devoret, M.H.},
  journal = {Phys. Rev. Appl.},
  volume = {11},
  issue = {5},
  pages = {054060},
  numpages = {19},
  year = {2019},
  month = {May},
  publisher = {American Physical Society},
  doi = {10.1103/PhysRevApplied.11.054060},
}

@article{Nilsson2023theory,
  title = {High-Gain Traveling-Wave Parametric Amplifier Based on Three-Wave Mixing},
  author = {Renberg Nilsson, Hampus and Fadavi Roudsari, Anita and Shiri, Daryoush and Delsing, Per and Shumeiko, Vitaly},
  journal = {Phys. Rev. Appl.},
  volume = {19},
  issue = {4},
  pages = {044056},
  numpages = {17},
  year = {2023},
  month = {Apr},
  publisher = {American Physical Society},
  doi = {10.1103/PhysRevApplied.19.044056},
}

@article{LeGal2025,
  title = {Gain compression in {Josephson} traveling-wave parametric amplifiers},
  author = {Le Gal, Gwenael and Butseraen, Guilliam and Ranadive, Arpit and Cappelli, Giulio and Fazliji, Bekim and Bonet, Edgar and Eyraud, Eric and Planat, Luca and Roch, Nicolas},
  journal = {Phys. Rev. Appl.},
  volume = {24},
  issue = {1},
  pages = {014022},
  numpages = {16},
  year = {2025},
  month = {Jul},
  publisher = {American Physical Society},
  doi = {10.1103/tq8k-m3dr},
}

@article{ranadive2025isolator,
  title={A travelling-wave parametric amplifier isolator},
  author={Ranadive, Arpit and Fazliji, Bekim and Le Gal, Gwenael and Cappelli, Giulio and Butseraen, Guilliam and Bonet, Edgar and Eyraud, Eric and B{\"o}hling, Sina and Planat, Luca and Metelmann, A and others},
  journal={Nature Electronics},
  volume={8},
  number={11},
  pages={1089--1098},
  year={2025},
  publisher={Nature Publishing Group UK London},
  doi={10.1038/s41928-025-01489-w}
}

@INPROCEEDINGS{Peng2022,
author = { Peng, Kaidong and Poore, Rick and Krantz, Philip and Root, David E. and O'Brien, Kevin P. },
booktitle = { 2022 IEEE International Conference on Quantum Computing and Engineering (QCE) },
title = {{ X-parameter based design and simulation of Josephson traveling-wave parametric amplifiers for quantum computing applications }},
year = {2022},
volume = {},
ISSN = {},
pages = {331-340},
doi = {10.1109/QCE53715.2022.00054},
publisher = {IEEE Computer Society},
address = {Los Alamitos, CA, USA},
month =sep}

@article{peng2022floquet,
  title={Floquet-mode traveling-wave parametric amplifiers},
  author={Peng, Kaidong and Naghiloo, Mahdi and Wang, Jennifer and Cunningham, Gregory D and Ye, Yufeng and O’Brien, Kevin P},
  journal={PRX Quantum},
  volume={3},
  number={2},
  pages={020306},
  year={2022},
  publisher={APS},
  doi={10.1103/PRXQuantum.3.020306}
}

@article{howe2026kinetic,
  title = {Kinetic inductance traveling-wave parametric amplifiers near the quantum limit: Methodology and characterization},
  author = {Howe, L. and Giachero, A. and Vissers, M. and Campana, P. and Wheeler, J. and Gao, J. and Austermann, J. and Hubmayr, J. and Nucciotti, A. and Ullom, J.},
  journal = {Phys. Rev. Appl.},
  volume = {25},
  issue = {4},
  pages = {044027},
  numpages = {20},
  year = {2026},
  month = {Apr},
  publisher = {American Physical Society},
  doi = {10.1103/yg71-j2dn},
}

@article{faramarzi2024kinetic,
    author = {Faramarzi, Farzad and Stephenson, Ryan and Sypkens, Sasha and Eom, Byeong H. and LeDuc, Henry and Day, Peter},
    title = {A 4–8 GHz kinetic inductance traveling-wave parametric amplifier using four-wave mixing with near quantum-limited noise performance},
    journal = {APL Quantum},
    volume = {1},
    number = {3},
    pages = {036107},
    year = {2024},
    month = {07},
    issn = {2835-0103},
    doi = {10.1063/5.0208110},
}

@article{aumentado2020superconducting,
  title={Superconducting parametric amplifiers: The state of the art in {J}osephson parametric amplifiers},
  author={Aumentado, Jose},
  journal={IEEE Microwave Magazine},
  volume={21},
  number={8},
  pages={45--59},
  year={2020},
  publisher={IEEE},
  doi={10.1109/MMM.2020.2993476}
}

@article{fadavi2023three,
  title={Three-wave mixing traveling-wave parametric amplifier with periodic variation of the circuit parameters},
  author={Fadavi Roudsari, Anita and Shiri, Daryoush and Renberg Nilsson, Hampus and Tancredi, Giovanna and Osman, Amr and Svensson, Ida-Maria and Kudra, Marina and Rommel, Marcus and Bylander, Jonas and Shumeiko, Vitaly and others},
  journal={Applied Physics Letters},
  volume={122},
  number={5},
  year={2023},
  publisher={AIP Publishing},
  doi={10.1063/5.0127690}
}

@article{chaudhuri2017broadband,
  title={Broadband parametric amplifiers based on nonlinear kinetic inductance artificial transmission lines},
  author={Chaudhuri, Saptarshi and Li, Dale and Irwin, KD and Bockstiegel, Clint and Hubmayr, Johannes and Ullom, JN and Vissers, MR and Gao, Jiansong},
  journal={Applied Physics Letters},
  volume={110},
  number={15},
  year={2017},
  publisher={AIP Publishing},
  doi={10.1063/1.4980102}
}

@article{giachero2024kinetic,
  title={Kinetic inductance traveling wave amplifier designs for practical microwave readout applications},
  author={Giachero, A and Vissers, M and Wheeler, J and Howe, L and Gao, J and Austermann, J and Hubmayr, J and Nucciotti, A and Ullom, J},
  journal={Journal of Low Temperature Physics},
  volume={215},
  number={3},
  pages={152--160},
  year={2024},
  publisher={Springer},
  doi={10.1007/s10909-024-03078-1}
}

@article{vissers2016low,
  title={Low-noise kinetic inductance traveling-wave amplifier using three-wave mixing},
  author={Vissers, Michael R and Erickson, Robert P and Ku, H-S and Vale, Leila and Wu, Xian and Hilton, GC and Pappas, David P},
  journal={Applied Physics Letters},
  volume={108},
  number={1},
  year={2016},
  publisher={AIP Publishing},
  doi={10.1063/1.4937922}
}

@article{esposito2021perspective,
  title={Perspective on traveling wave microwave parametric amplifiers},
  author={Esposito, Martina and Ranadive, Arpit and Planat, Luca and Roch, Nicolas},
  journal={Applied Physics Letters},
  volume={119},
  number={12},
  year={2021},
  publisher={AIP Publishing},
  doi={10.1063/5.0064892}
}

@article{metelmann2022parametric,
  title={Parametric couplings in engineered quantum systems},
  author={Metelmann, Anja},
  journal={SciPost Phys. Lect. Notes},
  number={66},
  year={2023},
  doi={10.21468/SciPostPhysLectNotes.66}
}

\end{document}